\documentclass[aps,twocolumn,superscriptaddress,amsmath,amssymb,longbibliography,showkeys,floatfix]{revtex4-1}

\usepackage{graphicx}
\begin{document}

\title{The mode of host-parasite interaction shapes coevolutionary dynamics \\ and the fate of host cooperation}

\author{Benjamin J. Z. Quigley}
\thanks{These authors contributed equally to this study.}
\affiliation{Department of Zoology, University of Oxford, South Parks Road, Oxford OX1 3PS, UK}
\author{Diana \surname{Garc\'ia L\'opez}}
\thanks{These authors contributed equally to this study.}
\affiliation{School of Physics and Astronomy, University of Manchester, Manchester M13 9PL, UK}
\thanks{Authors for correspondence: sam.brown@ed.ac.uk, \\diana.garcia@manchester.ac.uk}
\author{Angus Buckling}
\affiliation{Biosciences, University of Exeter, Penryn, Cornwall TR10 9EX, UK}
\author{Alan J. McKane}
\affiliation{School of Physics and Astronomy, University of Manchester, Manchester M13 9PL, UK}
\author{Sam P. Brown}
\affiliation{Institute of Evolutionary Biology, School of Biological Sciences, University of Edinburgh, King's Buildings, Edinburgh EH9 3JT, UK}
\thanks{Authors for correspondence: sam.brown@ed.ac.uk, \\diana.garcia@manchester.ac.uk}

\vspace{2mm}

\begin{abstract}
Antagonistic coevolution between hosts and parasites can have a major impact on host population structures, and hence on the evolution of social traits. Using stochastic modelling techniques in the context of bacteria-virus interactions, we investigate the impact of coevolution across a continuum of host-parasite genetic specificity (specifically, where genotypes have the same infectivity/resistance ranges (matching alleles, MA) to highly variable ranges (gene-for-gene, GFG)) on population genetic structure, and on the social behaviour of the host. We find that host cooperation is more likely to be maintained towards the MA end of the continuum, as the more frequent bottlenecks associated with an MA-like interaction can prevent defector invasion, and can even allow migrant cooperators to invade populations of defectors.
\end{abstract}


\maketitle

\section{\label{sec:introduction} Introduction}

The maintenance of cooperation is an evolutionary conundrum: why invest resources if the benefits of the investment are returned to other individuals? A major line of explanation for the persistence of cooperation in the face of non-investing `cheats' is that the benefits generated by cooperators return preferentially to cooperative individuals, as a result of non-random population structure \cite{frank1998foundations, rousset2004genetic, taylor2006direct, traulsen2009stochastic, szabo2007evolutionary, nowak2006evolutionary}. Here, we extend the study of cooperation by turning our focus to a ubiquitous driver of population
structuring: parasites.

Parasites shape host populations owing to the deleterious effects they have on their hosts \cite{anderson1979population, prado2009coevolutionary}. Given that parasites rely on their hosts for resources, changes in the host population composition will also impact upon parasite demography and genetic structure \cite{lion2010parasites}. A tight genetic interaction between hosts and pathogens can lead to ongoing host-parasite coevolution, defined as the reciprocal evolution of interacting hosts and parasites \cite{janzen1980coevolution}. The interaction is usually antagonistic, and selects for hosts to evolve resistance and parasites to evolve infectivity. A key consequence of coevolution is the impact on genetic diversity of host and parasite populations, although this will critically depend on the type of coevolution dynamics. At one extreme, coevolution follows an arms race dynamics (ARD), with directional selection for increasing resistance and infectivity, purging diversity. At the other is fluctuating selection dynamics (FSD), where parasite genotypes specialize on host genotypes, potentially resulting in the maintenance of large amounts of host and parasite genetic diversity in time and space \cite{frank1993specificity, gandon1996local, agrawal2002infection}.

An important determinant of coevolutionary dynamics is the underlying specificity. The two most common representations of specificity are the matching alleles (MA) model and the gene-for-gene (GFG) model, although other variants exist, for example \cite{fenton2009inverse}. MA models are based upon a system of self/non-self recognition molecules where hosts can successfully defend against any parasite genotype that does not match their own \cite{hamilton1980sex, grosberg2000mate}. Typical of many invertebrate immune systems, MA models assume that one parasite genotype will have a different subset of susceptible hosts than another parasite genotype. Infection is therefore determined by both the host and parasite genotypes, with such tight specificity sometimes leading to FSD \cite{salathe2008state}. The GFG model, favoured by plant pathologists, predicts whether infection is successful based on the interaction between resistance loci and virulence loci \cite{agrawal2002infection, parker1994pathogens, flor1956complementary}. GFG models are often characterized by directional ARD \cite{parker1994pathogens, thompson1992gene, sasaki2000host}, which can lead to the evolution of generalist parasites \cite{buckling2002antagonistic}, although if there are costs associated with increased resistance/infectivity ranges, FSD can also arise \cite{agrawal2002infection,sasaki2000host}.

As with coevolutionary dynamics themselves, MA and GFG models can be understood as the two extremes of a continuum of specificity, and the interactions between most hosts and parasites are likely to lie somewhere between the two extremes with some degree of specialization and some generalization. Agrawal \& Lively \cite{agrawal2002infection} investigated the dynamics of hybrid MA-GFG models via the use of a single parameter which formed a continuum: MA at one extreme and GFG at the other. Accurate characterizations of specificity patterns have important consequences for predicting the evolution of virulence \cite{kirchner2002evolutionary, dybdahl2003parasite}, patterns in local adaptation \cite{gandon2002local, morgan2005effect, nuismer2006parasite} and the evolution of recombination \cite{parker1994pathogens, otto2004species}. Now we ask: how does genetic specificity shape host-parasite coevolution, host population genetic structuring and in turn, the maintenance of host cooperation?

A recent study on cooperation in a bacterial model system highlighted that strong selection on the host population (adaptation to the passaging environment and to an antagonistic phage virus) allowed a primarily cooperative host population to purge low-frequency cheats, presumably owing to cooperator alleles hitchhiking on beneficial resistance mutations \cite{morgan2012selection}. Using a fully stochastic model of host-parasite dynamics, we dissect this bacteria-phage interaction, and extend to a coevolutionary time-scale, spanning GFG and MA mechanisms of interaction. We predict that the MA limit is the most favourable for the maintenance of cooperation because of increased genetic turnover allowing repeated purges of rare defector alleles.

\section{\label{sec:material_and_methods} Material and methods}

We follow a stochastic framework widely used in population dynamics \cite{mckane2004stochastic}. Numerical simulations were performed using an exact method, the Gillespie algorithm \cite{gillespie1977exact}. The analysis takes place within the framework of an ecological model of the infection of bacteria by lytic phages, using stochastic population dynamics. The parameter values used are based on biologically realistic values \cite{depaepe2006viruses}.

\subsection{\label{sec:simple_hostpar_model} Simple host-parasite model}

To start with, we consider a single type of host (a bacterium) and its parasite (a virus -- in particular, a lytic phage). The changes in the discrete numbers of bacterial hosts $n_H$ and free-living viruses $n_V$ are due to four processes: host birth, competition among hosts, virus death and lysis of a bacterium by a phage (resulting in the instantaneous death of the bacterium and the release of $y$ copies of the virus; we assume that the latent period is negligible). They can be captured using the following reactions:
\begin{equation}
H \xrightarrow{b} 2 H\,, \quad 2 H \xrightarrow{c} H\,, \quad V \xrightarrow{d} \emptyset\,, \quad  H + V \xrightarrow{p} y V\,.
\label{reactions_hp}
\end{equation}

From these microscopic reaction rates, we can derive equations for the average concentrations of hosts and parasites (see supplementary material S1), $H$ and $V$, which follow the deterministic trajectories (in a mean-field approximation):
\begin{eqnarray}
\frac{d H}{dt}&=& b\,H\left(1-\frac{H}{b/c}\right)- p\,VH \nonumber \\
\frac{d V}{dt}&=& p\,(y-1)\,VH -d\,V\,.
\label{det_equations_hp}
\end{eqnarray}

The system is similar to a predator-prey system \cite{mckane2005predator}, differing solely in the interpretation and magnitude of the quantity $y$ which stands for the \emph{burst size} and is of order $10$-$100$ -- while in a predator-prey system the quantity $(y-1)$ would be termed \emph{ecological efficiency} and takes a value less than $1$.

\subsection{\label{sec:hospar_coev_model} Host-parasite coevolution model}

The simplest scenario that allows for host-parasite coevolution is that between two types of host and two types of parasite. Using the notation of a single locus and two alleles, we will call the hosts $H_a$ and $H_A$ and allow for mutations among them, and similarly for the two types of parasite $V_a$ and $V_A$.

Depending on the model of specificity, the infectivity and susceptibility ranges of host and parasite (i.e. the parameters of the four different lysis reactions that can in principle take place) will vary. As in Agrawal \& Lively \cite{agrawal2002infection}, we will use a single parameter $q$ to translate across the MA-GFG continuum (MA: $q = 0$, GFG: $q = 1$). The parameter $p$ represents the binding efficiency of a specialist virus (which can only bind to one host type) and $p_g$ ($\leq p$) the binding efficiency of a generalist virus (which can bind to two different host types). We have chosen the form of these interactions to be a linear interpolation from MA to GFG, as summarized in table \ref{tab:lysis_rates}.

\begin{table}[h]
\begin{tabular}{|c|ccc|} \hline
	& $H_a$		& & $H_A$			\\ \hline
$\qquad V_a\qquad $	& $p\,(\equiv p_1)$		& & $0$			\\
$V_A$	& $q\,p_g\,(\equiv p_3)$	& $\quad$ & \small{$q\,p_g+(1-q)\,p\,(\equiv p_4)$} \\ \hline
\end{tabular}
\caption[]{\label{tab:lysis_rates} Interaction parameters (lysis rates) for the four host-parasite pairs of the coevolution model (\ref{sec:hospar_coev_model}).}
\end{table}

When $q = 0$ (at the MA end of the continuum), there are two specialist host-parasite pairs: $(V_a, H_a)$ and $(V_A, H_A)$. When $q = 1$ (at the GFG end), one virus is a specialist ($V_a$ can only infect $H_a$) but the other one is a generalist ($V_A$ can infect both $H_a$ and $H_A$), albeit at the cost of a smaller binding efficiency. From the point of view of the hosts, $H_A$ carries a resistance allele (to $V_a$) and incurs in a cost of resistance in the form of a death rate $z$ that scales with $q$. Therefore, the reactions that we must add to the ones considered in the simple host-parasite model are
\begin{eqnarray}
& & H_a \overset{\mu_H}{\longleftrightarrow} H_A, \ \ V_a  \overset{\mu_V}{\longleftrightarrow} V_A, \ \ H_A \xrightarrow{q z} \emptyset\,, \nonumber \\
& & V_a + H_a \xrightarrow{p_1} y V_a, \ \ V_A + H_a \xrightarrow{p_3} y V_A, \nonumber \\
& & V_A + H_A \xrightarrow{p_4} y V_A \,.
\label{reactions_coev}
\end{eqnarray}

Starting from the stochastic formulation and proceeding as in the previous simple host-parasite model, we can derive the following deterministic equations for the evolution of the average concentrations of hosts and parasites (see supplementary material S1):
\begin{eqnarray}
\frac{d H_a}{d t}&=& b H_a\left(1-\frac{H_a+H_A}{b/c}\right) +\mu_H(H_A - H_a) \nonumber\\
 & & -H_a(p_1 V_a +p_3 V_A)  \nonumber\\
\frac{d H_A}{dt}&=& b H_A\left(1-\frac{H_a+H_A}{b/c}\right) -qz H_A \nonumber\\
 & & -\mu_H(H_A - H_a) -p_4 V_A H_A \nonumber \\
\frac{d V_a}{dt}&=& p_1 (y-1) V_a H_a +\mu_V( V_A - V_a) - d V_a  \nonumber \\
\frac{d V_A}{dt}&=& (y-1) V_A (p_3 H_a + p_4 H_A) \nonumber\\
 & & -\mu_V(V_A - V_a) - d V_A \,.
\label{det_equations_coev}
\end{eqnarray}

Note that there is no direct competition between parasites, only an indirect one due to sharing common \emph{prey} (hosts) when $q > 0$. The effect of $q$ can be interpreted as a modification of the structure of the host-parasite network of interactions (see supplementary material S6). This coevolution model can in fact be taken to be a minimal motif within a generic host-parasite food web. The change from $q = 0$ (MA) to $q = 1$ (GFG) corresponds to the change from a modular to a nested network structure (diagonal and lower-triangular interaction matrices, respectively). For a recent study about the structure of experimental bacteria-phage networks, see Flores \textit{et al.} \cite{flores2011statistical}.

In our model, there is no clear time-scale separation between the ecological and the evolutionary dynamics (between the dynamics \textit{on} and \textit{of} the host-parasite network). Evolutionary dynamics here refers to the mutations allowed within the definition of the model (not to the evolution of the strength of the host-parasite links i.e. of $q$ itself -- that is left for future work). The stochastic population dynamics will take the ecosystem through a series of network configurations, as nodes are populated and links activated -- due to mutations -- or removed -- due to extinctions. For an example of a model of community assembly and for further references, see Capit\'an \& Cuesta \cite{capitan2011species}.

\subsection{\label{sec:coevcoop_model} Cooperation and coevolution model}

We introduce a third model by adding a cooperation- defection dilemma to the coevolution model described in \ref{sec:hospar_coev_model}. Now every host has two traits: the original one (that determines resistance to parasites) labelled by $a$ and $A$, plus a social trait labelled $C$ for cooperators and $D$ for defectors. So we consider four types of host in total: $C_a$, $D_a$, $C_A$ and $D_A$, as well as the two types of viruses $V_a$ and $V_A$.

We capture the social dilemma posed by the interaction between cooperators and defectors by cooperators paying a fixed cost ($w$), while providing benefits to every individual in the local population, be they a cooperator or a defector. Specifically, we include an extra birth rate for all hosts, proportional to the fraction of cooperators in the population (see supplementary material S1). The reactions we must add are the cost of cooperation and mutations in the cooperation-defection trait:
\begin{equation}
C_a \xrightarrow{w} \emptyset\,, \quad C_A \xrightarrow{w} \emptyset\,, \quad C_a \overset{\mu_{cd}}{\longleftrightarrow} D_a\,, \quad C_A \overset{\mu_{cd}}{\longleftrightarrow} D_A\,.
\label{reactions_coop}
\end{equation}

\section{\label{sec:results_and_discussion} Results and discussion}

To analyse the influence of host-parasite coevolution on the maintenance of host cooperation, we take a step-wise approach. First, we consider the impact of host evolution on host genetic structure, in the face of a static environmental challenge (following Morgan \textit{et al.} \cite{morgan2012selection}). We then introduce coevolutionary dynamics, without explicit cooperation and defection, to explore the basic impacts on host structure of reciprocal evolutionary antagonism. Finally, we analyse the full model, to track the fate of cooperation in the context of antagonistic coevolution.

\subsection{\label{sec:res_host_evol_only} Host evolution only}

\begin{figure}
\includegraphics[trim=0mm 0mm 0mm 0mm,clip,angle=0,width=0.48\textwidth]{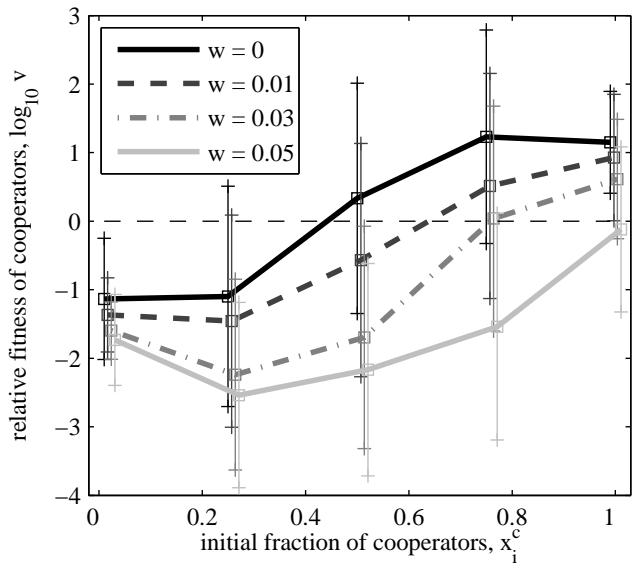}
\caption[]{\label{fig:1} Relative fitness of cooperators across different starting frequencies, for increasing cost of cooperation $w$ (see \ref{sec:res_host_evol_only} and supplementary material S2). Cooperators demonstrate positive frequency-dependent selection. The curve is shifted to the right as the cost of cooperation increases, i.e. higher initial frequencies of cooperators are required to maintain cooperation. Parameter values: $b_1 = 0.2\,b_0$ (except in the $w = 0$ curve, where $b_1 = 0$), $q = 0$, $\mu_V = 0$, $\mu_{cd} = \mu_H = 10^{-5}$ h$^{-1}$, $b_0 = 1$ h$^{-1}$, $K = b_0/c = 10^{6}$ ml$^{-1}$, $p = 10^{-6}$ ml h$^{-1}$, $y = 50$, $d = 0.2$ h$^{-1}$, volume $= 1$ ml, final time $t_f = 100$ h, statistics over $200$ runs per parameter set. Initial conditions: $V_a = 10^6$ ml$^{-1}$, $C_a + D_a = 10^3$ ml$^{-1}$, $V_A = C_A = D_A = 0$.}
\end{figure}

We begin by simplifying the full model (\ref{sec:coevcoop_model}) by removing parasite evolution (starting with $V_a$ only and setting $\mu_V = 0$), in order to mimic the system studied in Morgan \textit{et al.} \cite{morgan2012selection}. In agreement with Morgan \textit{et al.} \cite{morgan2012selection}, we find that strong selection on a non-social trait (here parasite resistance) generates positive frequency-dependent selection for cooperation (see figure \ref{fig:1} and supplementary material S2). The simple heuristic model in Morgan \textit{et al.} \cite{morgan2012selection} offered an interpretation for this effect: because beneficial non-social mutations are more likely to arise in the numerically dominant cooperator population, the cooperation allele is likely to hitchhike with these beneficial mutations that sweep through the population. However, the model paid for its simplicity with a number of stringent assumptions: no cost of cooperation, no demographic dynamics, deterministic gene frequency change and no coevolution.

Introducing a cost of cooperation in our explicitly dynamical and stochastic framework illustrated that, as the relative cost of cooperation increases, cooperators are required to be initially present at higher frequencies in order for cooperation to be maintained (figure \ref{fig:1}). When there is no cost of cooperation, the positive frequency-dependent effect of strong selection favours the maintenance of the more common genotype (more than $50\%$). As the cost of cooperation is increased, this `break-even' threshold is shifted towards higher frequencies of cooperators, as the intrinsic cheater advantage to defectors becomes increasingly important. This result aids the interpretation of Morgan \textit{et al.}'s experimental results, where the break-even point ranged from about $90$ per cent cooperators (environment and phage adaptation) to about $98$ per cent cooperators (environment adaptation only). These biases from the purely symmetrical `survival of the commonest' result (figure \ref{fig:1}, $w = 0$) reflect an increasing cost of cooperation, relative to the presumed benefits of adaptation to the environmental challenge.

\begin{figure}
\includegraphics[trim=0mm 0mm 0mm 0mm,clip,angle=0,width=0.48\textwidth]{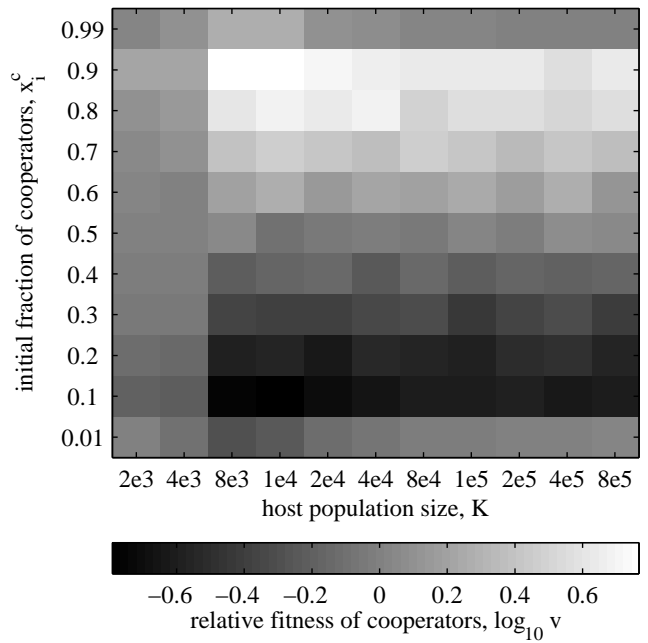}
\caption[]{\label{fig:2} Average relative fitness of cooperators with respect to defectors, measured by the quantity $\log_{10} v$ (see \ref{sec:res_host_evol_only} and supplementary material S2), as a function of the initial fraction of cooperators and of the host population size. The frequency-dependent effect depicted in figure \ref{fig:1} breaks down at low densities. In this scenario, the virus does not evolve -- but hosts may mutate to a resistant form. Parameter values are $b_1 = 0$, $w = 0$, $\mu_{cd} = \mu_H = 10^{-4}$ h$^{-1}$, $1000$ runs per parameter set; other parameters and initial conditions as in figure \ref{fig:1}. }
\end{figure}

The Morgan \textit{et al.} \cite{morgan2012selection} model predicted that the frequency- dependent effect highlighted in figure \ref{fig:1} would break down at low densities (as neither cooperator nor defector lineages are likely to gain resistance) and also at high densities (as both are likely to gain resistance, again removing the relative advantage to either lineage). In figure \ref{fig:2}, we revisit these predictions, and find in contrast that the frequency-dependent effect is robustly maintained at higher densities. Our explicitly dynamical and stochastic model allows resistance mutations to be separated in time, so even if at high densities both lineages are likely to gain a critical resistance mutation, there is still a decisive advantage to gaining the resistance mutation first -- and this is more likely to be the case in the dominant (higher frequency) lineage.

\subsection{\label{sec:res_hostpar_coev} Host-parasite coevolution}

We now turn our focus to the effects of coevolution on the host population dynamics, first in the absence of any social conflict (as described in \ref{sec:hospar_coev_model}). Note that the reciprocally antagonistic nature of the bacteria-virus interaction can readily lead to instability, and loss of the pairing in a spatially unstructured population \cite{schrag1996host}. In order for the system to be stable (i.e. for hosts and parasites to coexist, without global extinctions), a substantial cost of generalism must be imposed (see supplementary material S4), both on the resistant host and on the generalist parasite. A cost of wider viral infectivity range is expressed in our model in terms of a reduced binding efficiency, consistent with experimental data \cite{poullain2008evolution} and a trade-off mechanism based on antagonistic pleiotropy \cite{elena2009evolutionary}, i.e. mutations that allow the virus to infect a broader host range trade off against viral productivity. A cost of generalized resistance is captured in our model via an intrinsic per capita cost to the bacterial host, consistent with experimental data \cite{ferdig1993reproductive, boots1993trade, fellowes1998trade, langand1998cost, hall2011host}.

\begin{figure}
\includegraphics[trim=0mm 0mm 0mm 0mm,clip,angle=0,width=0.48\textwidth]{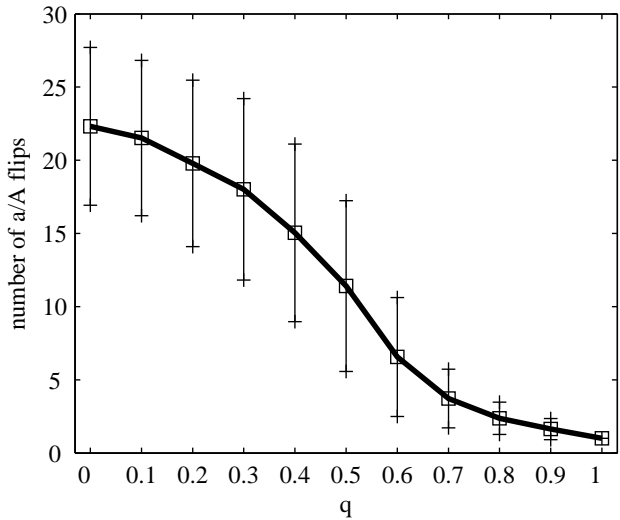}
\caption[]{\label{fig:3} Host genotype substitution dynamics along the MA-GFG continuum in the host-parasite coevolution model (see \ref{sec:hospar_coev_model} and \ref{sec:res_hostpar_coev}). Number of $a/A$ flips in a $1000$ h interval; we consider that there is a flip every time that the fraction of type $a$ hosts drops below $0.95$. Averages and $\pm 1$ s.d. over $500$ runs are shown. Parameter values: $z = 0$, $b = 1$ h$^{-1}$, $K = b/c = 10^6$ ml$^{-1}$, $\mu_H = \mu_V = 10^{-5}$ h$^{-1}$, $p = 10^{-6}$ ml h$^{-1}$, $p_g = 0.1\,p$, $y = 30$, $d = 0.2$ h$^{-1}$, volume $= 1$ ml. Initial conditions: $V_a = 10^6$ ml$^{-1}$, $H_a = 10^2$ ml$^{-1}$, $V_A = H_A = 0$.}
\end{figure}

Following the establishment of a robust coexistence regime (see supplementary material figure S4), we turn to an analysis of host population bottlenecks as a function of the specificity of the genetic interaction between host and parasite. Consistent with Agrawal \& Lively \cite{agrawal2002infection}, we observe a reduced number of host population bottlenecks as $q$ is increased -- i.e. moving from MA to GFG (figure \ref{fig:3}). Underlying each bottleneck is in fact an attempt of invasion by the $A$ genotype (originated by a mutation) into the resident $a$ population. The structure of the host-parasite network of interactions, fixed by the value of $q$ (see supplementary material S6), determines the outcome. Invasion attempts fail repeatedly at the MA end, as the invading $H_A$ genotype causes an explosion in its specialist opponent $V_A$, which in turn abruptly depletes the $H_A$ population in a negative feedback loop (so we observe many bottlenecks, as the $a/A$ battle continues).

In contrast, a positive value of $q$ stabilizes the system, preserving a balanced polymorphism in host resistance alleles. Indeed, at the GFG end $H_A$ manages to successfully invade once and then coexist with the resident $H_a$ due to the extra link in the network (supplementary material S6, its virus $V_A$ is devoting part of its resources to attack $H_a$ too) and because the costs of generalism come into play: the binding efficiency of $V_A$ to $H_A$ is lower; the intrinsic growth rate of $H_A$ is also lower due to the cost of generalism, paradoxically raising its chances of a successful invasion. It seems that these costs of generalism increase the period of the $H_A-V_A$ host-parasite oscillation, thus enabling the successful establishment of $H_A$ (see supplementary material S7).

\subsection{\label{sec:res_coevcoop} Coevolution and cooperation}

We now study the full model of \ref{sec:coevcoop_model}, where there is a cooperation-defection dilemma for the hosts, as well as host-parasite coevolution. We can now ask: what is the fate of cooperation as we move along the host-parasite specificity continuum from the MA ($q = 0$) to the GFG ($q = 1$) limits?

\begin{figure}
\includegraphics[trim=0mm 0mm 0mm 0mm,clip,angle=0,width=0.48\textwidth]{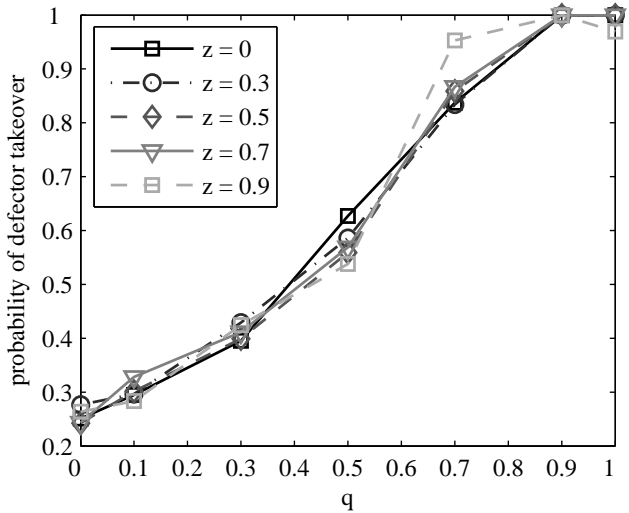}
\caption[]{\label{fig:4} Probability of defector takeover across the MA-GFG continuum (MA $q = 0$, GFG $q = 1$) in the full coevolution and cooperation model (see \ref{sec:coevcoop_model} and \ref{sec:res_coevcoop}) for different costs of generalism $z$. We plot the fraction of runs where defectors grow to be more than $99$ per cent of the initially cooperator-only host population, within a $2000$ h time window of observation and subject to non-extinction of the system (more than $85\%$ of runs are stable in this case). Parameter values are $p_g = 0.01\,p$, $w = 0.2$ h$^{-1}$, $K = b_0/c = 10^6$ ml$^{-1}$, $b_1 = 0.2\,b_0$, $\mu_V = \mu_H = \mu_{cd} = 10^{-5}$ h$^{-1}$, $1000$ runs per set, other parameters as in figure \ref{fig:2}. Initial conditions: $V_a = 10^6$ ml$^{-1}$, $C_a = 10^2$ ml$^{-1}$, $V_A = C_A = D_a = D_A = 0$. Cooperation is maintained for longer times under a MA model of specificity.}
\end{figure}

The MA model is associated with a diversification of both host and parasite genotypes \cite{woolhouse2000search} and so at first sight is the most inimical to the preservation of cooperation, which is typically favoured by local genetic homogeneity \cite{west2007evolutionary}. However, we find that cooperation is most robust to cheater invasion in the MA limit (figure \ref{fig:4}). The probability of defector takeover grows with $q$ as we move from MA to GFG (figure \ref{fig:4}). In other words, cooperation is maintained for longer periods if host-parasite specificity is MA-like. This result is robust across a large region of parameter space (although system stability may be limited -- see \ref{sec:res_hostpar_coev}).

The non-intuitive association between diversifying MA-type interactions and the preservation of cooperation can be understood when it is recognized that the diversifying selection is not acting on the cooperation locus, but only on the resistance locus. The result of the continual strong alternating selection on resistance is a continual cycle of genetic bottlenecks purging diversity at linked loci, which here means the cooperation locus. In the MA limit, recurrent $a/A$ bottlenecks prevent rare defectors from invading -- so cooperation is maintained (any $D_A$ defectors that may have arisen are wiped out, just like the majority of $C_A$, due to the burst in $V_A$). On the contrary, in the GFG limit, a single switch from $a$ to $A$ hosts paves the way for subsequent defector takeover on a static $A$ allele background. Examples of the corresponding typical time series are shown in supplementary material figure S3.

We show that the genetic structure of host-parasite interaction influences the outcome of their coevolutionary dynamics. These dynamics in turn determine the structuring of the host population via genetic bottlenecks, which we find to be more numerous towards the MA end of the MA-GFG continuum. Population bottlenecks are known to influence the social behaviour of the host \cite{brockhurst2007population}, favouring cooperation.

\subsection{\label{sec:res_spatial_structure} Spatial structure}

The assumption that individuals encounter one another at random in well-mixed populations is made primarily for mathematical simplicity and tractability \cite{tilman1997spatial, dieckmann2000geometry}. In natural populations, however, this assumption is misplaced. Biological populations are largely broken up into subpopulations linked loosely by migration \cite{hanski2004ecology, gross2006metacommunities}, known as a metapopulation. Within a single patch, cheats have a selective advantage and cooperation inevitably breaks down \cite{west2003cooperation, griffin2004cooperation, chuang2009prisoner}. However, in the context of a metapopulation, cooperating demes grow to higher densities than cheating demes, enabling cooperation to persist across a spatially structured population \cite{west2003cooperation, griffin2004cooperation, chuang2009prisoner}.

In order to interpret our results in a metapopulation context, we explore how the dynamics within a patch is altered after a single migration event in a host-evolution- only scenario (as in \ref{sec:res_host_evol_only}). Each combination of arriving and resident genotype is tested, for different values of the migrant-to-resident ratio and of the time of arrival. See supplementary material S5 for details and figures. In contrast to the classical invasion hierarchy between cooperators and defectors, we find that rare cooperators can successfully invade locally abundant defectors if they carry the favourable resistance allele. Of course, the converse also holds, defectors can readily invade cooperators if they carry an additional resistance advantage (see supplementary material figure S5). If, however, both are susceptible, resident cooperators have a high chance of resisting an invasion (within our observation time window) unless the migrant defectors are very numerous. Regarding relative timings, the later the resistant migrants arrive, the higher the chance that the residents have already evolved a resistant mutation themselves -- which neutralizes the competitive advantage of the migrants.

A complete study involving repeated migration and coevolution in a full metapopulation setting is outside the scope of this article. We have, however, examined the two key issues, invasion hierarchies and the longevity of cooperator patches. In figure \ref{fig:4}, we demonstrate how the combination of interaction genetics and coevolution can define the longevity of cooperator patches, in the face of defector challenge. Specifically, we find that in the MA limit, defector invasion is much impaired and so cooperator patches are able to persist for longer, increasing the global prevalence of cooperation across a metapopulation. Expanding our theoretical framework to incorporate more complex mechanisms of interaction (e.g. inverse GFG, multi-locus interactions, \cite{fenton2009inverse, sasaki2000host}) and non-random host-parasite interactions will present important challenges for future work. Alongside theoretical development, further experimental tests are vital, and also offer important applied considerations.

\subsection{\label{sec:res_applied_context} Applied context}

Our theoretical analysis has largely been grounded in the biology of bacteria-phage interactions \cite{morgan2012selection}. This interaction has enormous public health interest \cite{woolhouse2002biological}, due to the clinical importance of bacterial pathogens and the increasing interest in the use of `phage therapy' as a novel mechanism of pathogen control \cite{sulakvelidze2001bacteriophage, skurnik2006phage}. Supplementary material figure S4 highlights a concern for phage therapy -- the administration of a lethal phage virus does not always result in the eradication of the target bacterial population, even in the absence of spatial structure. For much of the parameter range, the bacterial population evolves resistance to the phage and continues to persist, despite the ability of the phage to co-evolve and chase the bacteria through genotype space.

We go on to show that the nature of this coevolutionary persistence has great significance for the fate of bacterial cooperation, defining the ability of cooperators to resist local replacement by cheats. Bacterial cooperation typically involves the secretion of shared extracellular enzymes, toxins and polymers -- important virulence factors when released within a human host \cite{west2007evolutionary, nogueira2009horizontal}. Therefore, the coevolutionary interaction with phages may have an important impact on the virulence of bacterial pathogens. In the MA limit, diversifying selection on bacterial resistance alleles will pose a formidable barrier to the invasion of cheats from rare, increasing the resistance of resident wild-type bacteria to recently proposed mechanisms of pathogen control via the introduction of `Trojan Horse' cheater lineages \cite{harrison2006cooperation, brown2009social}.

A recent study has demonstrated that coevolutionary arms races between bacteria and phage decelerate over time \cite{hall2011host}, giving way to fluctuating selection. Moreover, bacteria-phage coevolution within natural soil communities follows an FSD \cite{ gomez2011bacteria}. These results suggests that MA-type dynamics, characteristic of fluctuating selection, are likely to be important in determining the phenotypic properties of parasite and host populations. Given the positive effect on cooperation at the MA extreme, this highlights the potential of our model to help explain the maintenance of cooperation in natural populations in a broader context.

\begin{acknowledgments}
We thank EPSRC grant EP/H032436/1, BBSRC, and the Wellcome Trust, grant ref WT082273 for funding. A.B. is supported by the ERC. We also thank two anonymous reviewers for their helpful and insightful comments.
\end{acknowledgments}

\bibliography{coevcoop}

\end{document}